\documentclass[12pt,preprint]{aastex}

\newcommand{\ggt}{GG\,Tau}
\newcommand{\mic}{\,$\mu$m}

\shorttitle{A multi-wavelength scattered light analysis of the
GG\,Tau ring} 
\shortauthors{Duch\^ene et al.}

\received{2003 December 13}
\begin{document}
\title{A multi-wavelength scattered light analysis of the
dust grain population in the GG\,Tau circumbinary ring}

\author{G. Duch\^ene, C. McCabe, A. M. Ghez}
\affil{Division of Astronomy and Astrophysics, UCLA, Los Angeles, CA
90095-1562}
\email{duchene@astro.ucla.edu}
\author{B. A. Macintosh}
\affil{Lawrence Livermore National Laboratory, 7000 East
Avenue, Livermore, CA 94550-9234}

\begin{abstract}
We present the first 3.8\mic\ image of the
dusty ring surrounding the young binary system \ggt, obtained
with the W. M. Keck\,II 10\,m telescope's adaptive optics
system. This is the longest wavelength at which the ring has
been detected in scattered light so far, allowing a
multi-wavelength analysis of the scattering properties of the
dust grains present in this protoplanetary disk in combination
with previous, shorter wavelengths, $HST$ images. We find that
the scattering phase function of the dust grains in the disk
is only weakly dependent on the wavelength. This is inconsistent
with dust models inferred from observations of the
interstellar medium or dense molecular clouds. In
particular, the strongly forward-throwing scattering phase
function observed at 3.8\mic\ implies a significant increase
in the population of large ($\gtrsim1$\mic) grains, which
provides direct evidence for grain growth in the
ring. However, the grain size distribution required to match
the 3.8\mic\ image of the ring is incompatible with its
published 1\mic\ polarization map, implying that the dust
population is not uniform throughout the ring. We also show
that our 3.8\mic\ scattered light image probes a deeper layer
of the ring than previous shorter wavelength images, as 
demonstrated by a shift in the location of the inner edge of 
the disk's scattered light distribution between 1 and 3.8\mic. We
therefore propose a stratified structure for the ring in which
the surface layers, located $\sim50$\,AU above the ring
midplane, contain dust grains that are very similar to those
found in dense molecular clouds, while the region of the ring located
$\sim25$\,AU from the midplane
contains significantly larger grains. This stratified structure 
is likely the result of vertical dust settling and/or preferred
grain growth in the densest parts of the ring.
\end{abstract}


%________________________________________________________________

\section{Introduction}

The process of planet formation is thought to occur during the
first few million years of the evolution of a star,
i.e. during its pre-main-sequence phase for low-mass
objects. The generally accepted scenario of planet formation,
as articulated originally by Safranov (1969), starts with an initial
population of small, interstellar medium (ISM)-like particles growing
by coagulation to form km-sized bodies, whose gravitational well
is then deep enough to ignite a runaway accretion process of
all dust and gas located at similar distances from the central
object. Numerical models of dust coagulation suggest that submicron
grains grow to
particles several times 10\mic\ or more in radius in less
than 10$^4$\,yrs in the densest parts of protoplanetary disks
(e.g, Weidenschilling 1980; Suttner \& Yorke
2001). Coagulation of small planetesimals into larger bodies
may then occur on timescales as short as $10^5$\,yrs (e.g.,
Wetherill \& Stewart 1993), with planetary cores ($R>10^3$\,km)
eventually forming within a
few AU of the central star over a timescale of $\lesssim10$\,Myr
(e.g., Pollack et al. 1996; Kenyon \& Bromley 2003). Observationally,
however, the timescales associated with each of these
stages of the planet formation process are poorly constrained. Using
the fraction of stars with a near-infrared excess within star-forming
regions of different ages as a proxy for disk evolution, Haisch, Lada
\& Lada (2001) estimate that the inner disk region is completely
cleared of \mic-sized particles in $\lesssim10$\,Myr. Whether this
timescale corresponds to the end of the runaway accretion phase of
planetesimals or to the disk dispersal through accretion onto the star
is unknown. In any event, to study grain growth during the
planet formation process, it is necessary to establish the properties
of dust grains in disks surrounding Myr-old T\,Tauri stars.

The search for empirical evidence of grain growth in
protoplanetary disks began through investigations that probed 
the (sub)millimeter thermal emission properties of the dust. 
In this regime,
the disks are largely optically thin, which allows for
relatively simple modeling. The large dust opacities
(Beckwith et al. 1990) and shallow spectral index of the
disk thermal emission (e.g., Beckwith \& Sargent 1991;
Mannings \& Emerson 1994) disagree with the expected behavior of
ISM-like dust grains and suggest that mm-sized grains are
present in most T\,Tauri circumstellar disks. Analyses of the
entire disk spectral energy distribution has recently led
to similar conclusions (D'Alessio, Calvet \& Hartmann 2001;
Wood et al. 2002). However, in order to devise an accurate model 
of the thermal emission from a circumstellar disk, the outer 
radius and vertical structure must Di determined from spatially 
resolved images; in the many
cases where this is not possible, degeneracies between the dust
temperature, total mass and opacity are left unsolved and cause 
large uncertainties in the size distribution (e.g.,
Chiang et ALU. 2001).

A second approach used to determining the properties of dust grains
in circumstellar disks is through high angular resolution
scattered light images in the visible or near-infrared. At these
wavelengths, a typical T\,Taro disk is optically 
thick; the photons, which are scattered off the surface layer of the
disk, therefore provide information on the global geometric properties
of the disk, such as its radius, total height and amount of
flaring. More importantly, the scattering process is dependent on the
size distribution of dust grains present in the disk. In the Mi
theory, which assumes scattering off spherical particles, the
scattering is a strong function of the dust grain ``size parameter''
($x=2\pi a/\lambda$, where $a$ is the grain radius), with 
scattered light images being most sensitive to grains in the range
$0.5\lesssim x\lesssim3$. Multi-wavelength scattered
light studies of disks therefore provide direct information on
the dust grain size distribution. For instance, scattered light
observations in the visible and near-infrared (1--2\mic) are sensitive
to grains that are, at most, 1\mic\ in size. Such observations alone
are therefore not enough to reach firm conclusions regarding grain
growth with respect to ISM particles, as the largest grains ISM grains
have inferred radii $\lesssim1$\mic\ (e.g., Mathis \& Whiffen 1989;
Kim, Martin \& Hendry 1994; Weingartner \& Draine 2001). 

Extending this multi-wavelength scattered light approach to
even longer wavelengths, i.e., in the thermal infrared regime
($>3$\mic), can allow one to probe the presence of grains
several microns in size, which are explicitly larger than
those found in the ISM and at least as large as those found in
dense molecular clouds (Weingartner \& Draine 2001). A
powerful approach to determining the dust grain size distribution,
therefore consists 
in obtaining several images of the same protoplanetary disk
over a wide range of wavelengths, ideally from the visible to
the mid-infrared. If this can be achieved, the
wavelength-dependence of the scattered light informs us
directly on the size distribution of dust grains as the disk
geometry is fixed and the detailed composition of the grains
only plays a limited role in their scattering properties. In
a recent study, we have obtained the first
11.8\mic\ scattered light image of the edge-on disk system
HK\,Tau\,B, which has clearly revealed the presence of dust
grains up to $\sim3$\mic\ (McCabe, Duch\^ene \& Ghez 2003).

We present here the first spatially-resolved 3.8\mic\ image of
the dusty torus surrounding the young (1\,Myr) \ggt\,A binary
system\footnote{\ggt\ is a hierarchical quadruple system, with
a 10\arcsec\ separation between \ggt\,A and
\ggt\,B. In the following, we only consider \ggt\,A, which we
refer to as \ggt\ for simplicity.}. This disk, first
discovered through its thermal emission by millimeter
interferometry (Simon \& Guilloteau 1992; Guilloteau, Dutrey
\& Simon 1999), is a massive (0.13\,$M_\odot$) torus with
inner and outer radii of 180 and 260\,AU located at a distance
of 140\,pc (Bertout, Robichon \& Arenou 1999). The inner
binary, which has a projected separation of 35\,AU (Leinert et
al. 1991; Ghez, Neugebauer \& Matthews 1993), has carved out a
central gap through its orbital motion, explaining the large
inner radius of the ring. Since its discovery, the ring has
been spatially resolved in scattered light at various
wavelengths ranging from 0.6 to 2\mic\ (Roddier et al. 1996;
Silber et al. 2000; McCabe, Duch\^ene \& Ghez 2002; Krist, Stapelfeldt
\& Watson 2002; Itoh et al. 2002). Models of the scattered light
images of the ring have both concluded for and against grain
growth in the disk (Wood, Crosas \& Ghez 1999; Krist et al. 2002;
McCabe et al. 2002), mostly because of their limited wavelength
coverage.

Our new 3.8\mic\ image of the \ggt\ torus clearly shows that
the deeper parts of the disk contain larger grains than the
surface layers, and that they are even larger than those found
in the ISM or in molecular clouds. The outline of this paper
is as follows: the observations and data reduction processes
are summarized in \S\,\ref{sec:obs}, while the results are
presented in \S\,\ref{sec:results}. Multi-wavelength Monte
Carlo models of the ring are presented in
\S\,\ref{sec:models}. The properties of dust grains in the
ring are discussed in \S\,\ref{sec:discus} and
\S\,\ref{sec:concl} summarizes our conclusions.

%________________________________________________________________

\section{Observations and data analysis}
\label{sec:obs}

The data presented here were obtained with the W. M. Keck\,II
10\,m telescope on 2002 December 12 \& 13, using the facility
NIRC2 instrument (K. Matthews et al., in prep.) installed
behind the adaptive optics (AO) system (Wizinowich et
al. 2000). With a 1024$^2$ detector and employing a pixel
scale of 0.00994\,\arcsec/pix, the instrument offers a field
of view of $\sim10$\arcsec. We used the $L'$ filter
($\lambda_0=3.78\,\mu$m, $\Delta\lambda=0.70\,\mu$m)
throughout the observations. A circular cold pupil mask,
equivalent to a 9\,m entrance pupil (largest circle inscribed
in the hegaxonal primary mirror), was used to reduce the
thermal background and smooth out the complex, hexagonal shape
of the telescope's point spread function (PSF). \ggt\ itself
($R=11.2$\,mag, Kenyon \& Hartmann 1995) was used as guide
star for the AO system, which was running at a frequency of
about 100\,Hz. The resulting AO correction was stable
throughout the run, with typical Strehl ratios on order of
60\,\% as measured on images of single stars relative to the
9\,m pupil employed for our observations. From a gaussian fit
to images of single stars, we estimate that the spatial
resolution of our dataset is about 0\farcs091 (FWHM).

The photometric standard HD\,1160 ($L'=7.04$\,mag, Bouchet,
Schmider \& Manfroid 1991; Tinney, Mould \& Reid 1993) was
observed on both nights of the observing run. Aperture
photometry was performed on this object with a circular
aperture of 1\farcs44 and measuring the sky level in an
annulus from 1\farcs49 to 1\farcs79. This allowed us to
estimate our absolute calibration scale to within
$\sigma_{L'}=0.13$\,mag. The airmass extinction was derived
through the repeated observations of CI\,Tau during the run
(see below) and we found an extinction coefficient of 0.048
mag/airmass on both nights. 

Short exposures (0.181\,s) of the \ggt\ system were obtained in
order to avoid saturation of the strong thermal background at
these wavelengths; 300 such exposures obtained at a given
position on the chip were coadded into a single image before 
nodding to another
position. We used a 4-position pattern defined so that the
whole ring would fit in the detector on each individual image
without overlap between consecutive images. Over the two
nights, we integrated for 32 such cycles on target, i.e. 128
independent images amounting to 6950\,s of integration
(1h56min). The images of \ggt\ were interleaved with
observations of blank sky fields and of isolated nearby stars
to estimate the instrumental PSF with a setup that reproduced
the same conditions for the AO system. We mostly used CI\,Tau
($R=12.2$\,mag, Kenyon \& Hartmann 1995), another T\,Tauri
star located $\sim5$\degr\ away from \ggt, and GJ\,9140
($R=11.3$\,mag, Weis 1996), both of which produced very
similar PSFs to the observations of \ggt. We held the
telescope's pupil orientation fixed throughout the
observations, so that the substructures in the wings of the
AO-corrected PSF remained stationary which allows for a
subsequent better subtraction. Consequently, the actual
orientation of the images on the sky continuously rotated at a
rate that never exceeded 2\degr\ per image and was
usually much slower, so that field rotation did not blur our 
image.

A nightly flat field image was created by medianing and
normalizing all images of the sky and of isolated standard
stars; 103 and 60 images were used in each night
respectively. In the images of single stars, all pixels with
fluxes higher than the 3\,$\sigma$ noise level were excluded
to avoid contamination of the stellar flux in the
flat-field. Each image of GG\,Tau was first sky-subtracted
using the image of the same cycle that i) had the object in
the opposite corner of the detector and ii) was closest in
time, and was then flat-fielded. Because the sky thermal
emission is rapidly variable, the sky subtraction did not
always yield a zero background. A secondary sky correction was
therefore performed by subtracting the median of the two star-
and disk-free quadrants from the whole image; this second
subtraction typically amounted to $\lesssim5\,\%$ of the
primary sky subtraction. Finally, the images were bad pixel
corrected.

To characterize the disk emission around \ggt, the much
stronger flux from the two components of the binary system had
to be removed. This was done through PSF fitting and
subtraction. We created a library of 5 PSF images of
CI\,Tau and/or GJ\,9140 for each night. Each image is the 
sky-subtracted
shift-and-add average of 1 to 3 consecutive 4-position cycles
on the same star. Since we combined only almost
contemporaneous images in each PSF image, each night's PSF
library contains a range of AO corrections, which allows for a
better fit of each image of \ggt. We created a PSF fitting
routine that extracts information for both components
(location and flux) simultaneously, using the built-in IDL
global minimization routine {\tt amoeba}, a downhill simplex
algorithm (Press et al. 1992), to minimize the residuals
within a 0\farcs5 radius around each component. Each of the
128 images of \ggt\ was PSF fitted using each PSF in the
library of the corresponding night; the final subtracted 
image used is that which yielded the smallest
residuals.

All images were then
rotated to a common orientation and shift-and-added using the
stars' location measured from the PSF fitting. This was done
separately on the datasets with and
without PSF subtraction; these final images are presented in
Figure\,\ref{fig:images}. Since the NIRC2 pixel scale
is 4.5 times smaller than the Nyquist sampling of the spatial
resolution of our images, we also applied a smoothing process
by replacing each pixel with the median value of a 3
pixel-radius aperture centered on it. This local averaging is
performed within less than one resolution element around each
pixel, so it does not affect our spatial resolution but it
significantly enhances the contrast of the images by
decreasing the pixel-to-pixel noise by a factor of $\sim$2.5.

The photometry and astrometry of the \ggt\ binary was
extracted from the PSF fitting results, using the average and
standard deviations of all measurements to estimate the final
values and their associated uncertainties. The flux for each
individual component is measured on the best-fitting model PSF
using aperture photometry within a 0\farcs50-radius aperture,
and measuring the sky in an annulus 2\farcs5--2\farcs6 in
radius in the actual image of the binary. The fluxes were
aperture-corrected: from encircled energy measurements of the
best-fitting PSF we find that a 0\farcs5 aperture contains
83\% of the total flux of any star. The results are summarized
in Table\,1. The measured flux ratio of the
binary ($\Delta L'=0.59\pm0.03$\,mag) is significantly
different from the $L'$ band ($\lambda_0=3.5\,\mu$m,
$\Delta\lambda=0.6\,\mu$m) flux ratio reported by White \&
Ghez (2001), $\Delta L=0.87\pm0.01$\,mag. This probably
results from the variability of at least one of the components
of the system, although the difference in filter bandpass
could account for part of this discrepancy. Regarding the
measured astrometric properties of the binary, we have taken
into account assumed uncertainties of 0.5\,\% on the
instrument plate scale measurement and 0\fdg5 on its absolute
orientation.

In the following, we adopt the stellar masses derived by White
et al. (1999), $M_A=0.76\,M_\odot$ and $M_B=0.68\,M_\odot$, to
determine the location of the center of mass of the binary
system. More recently, Hartigan \& Kenyon (2003) derived
$M_A=0.60\,M_\odot$ and $M_B=0.38\,M_\odot$, respectively,
using a more elaborate analysis that takes into account the
accretion-induced veiling in the stellar spectra. However,
their analysis of low- and medium- resolution spectra results
in discrepant estimates for the veiling, raising
questions about which set of stellar parameters to use. The
difference between the two predicted binary mass ratios,
$q=0.89$ from White et al. (1999) and $q=0.63$ from Hartigan
\& Kenyon (2003), results in an apparent shift of the center of
mass of the binary of about 0\farcs02 only, which is a
negligible effect for our analysis of the ring properties.

%________________________________________________________________

\section{Results}
\label{sec:results}

% - - - - - - - - - - - - - - - - - - - - - - - - - - - - - - - - 

\subsection{$L'$ properties of the circumbinary disk}
\label{subsec:ring_l}

Our PSF-subtracted $L'$ image of the \ggt\ system reveals a
continuous ring around the two stars with its brightest part
located at a position angle of $\sim25$\degr. The ring is
detected at most azimuths above the 2\,$\sigma$ detection
limit per resolution element, except for the sector at
position angles 140--180\degr, and up to $\sim10\,\sigma$ at
peak brightness.

To quantify the morphology and photometric properties of the
ring, we first fitted its structure with an ellipse. We used
the same approach as in McCabe et al. (2002), which we briefly
describe here. For each 10\degr\ sector in position angle, we
computed the radial profile of the ring, taking the center of
mass of the binary as reference. We then fitted a 4th-order
polynomial function to a $\sim$0\farcs5 portion of this
profile and extracted from this the location of the peak
(defined as the midpoint between the inner and outer half-peak
radii) and its intensity. The polynomial fit failed for the
sectors at position angles $\sim$100\degr\ and 140--200\degr\
due to the faintness of the ring in those areas. Uncertainties
in the peak position for each azimuthal profile are defined as
the standard deviation of the peak positions measured in four
sub-maps of the data. We then fitted an ellipse to the
location of the peaks at all available position angles with a
$\chi^2$ minimization routine. The best fitting model has a
semi-major axis of 1\farcs48$\pm$0\farcs02, an eccentricity of
0.65$\pm$0.01 and a position angle of the minor axis of
13\degr$\pm$5\degr. The geometrical center of the best fitting
ellipse is located 0\farcs16$\pm$0\farcs04 due South from the
center of mass of the binary.

Since the apparent morphology of the ring is in agreement with
that derived from other images of the ring (see
\S\,\ref{subsec:ring_all}), we estimated the photometry of the
ring through the same aperture as in McCabe et al. (2002),
defined by two concentric, $e=0.65$ ellipses of radii
0\farcs99 and 2\farcs2 respectively. The integrated flux of
the disk corresponds to 0.97\,\% of the integrated starlight
from both components (with a relative uncertainty on the order
of 10\,\%), and its peak surface brightness is about
11.3\,mag/arcsec$^2$. Using the same elliptical aperture, we
find that the flux ratio between the northern (position angles
275--95\degr) and southern (position angles 95--275\degr)
halves of the disk is $\sim$2.5. From the polynomial fits
presented above, we constructed an azimuthal intensity profile
of the ring (Figure\,\ref{fig:profilepa}), from which we find
that the peak-to-peak intensity is $\sim$5, although there are
a few sectors of the ring whose (undetected) maximum intensity
is $\gtrsim6$ times fainter than its peak surface
brightness. Since the southern half of the ring is much
fainter than the northern half, those flux ratios are subject
to 20--30\,\% uncertainties which are dominated by
uncertainties in the background subtraction. These results are
summarized in Table\,2.

% - - - - - - - - - - - - - - - - - - - - - - - - - - - - - - - - 

\subsection{Wavelength dependence of the ring properties}
\label{subsec:ring_all}

In this section, we compare the properties of the ring
observed in our $L'$ image to those obtained at shorter
wavelengths by Krist et al. (2002) and McCabe et
al. (2002). In spite of a significantly higher background
noise due to the atmosphere's thermal emission at 3.8\mic, our
new image is of high enough quality to allow us to measure
significant wavelength-dependent effects. Given the very low
temperature of the dust in the ring (35\,K, Guilloteau et al. 1999),
we can only be detecting scattered starlight off the ring at
3.8\mic. This is so far the longest wavelength at which this
structure has been detected in scattered light. Quantifying
the chromatic effects in the ring images will help us
constrain the properties of the dust population.

Table\,2 summarizes several quantities
extracted from our $L'$ image, as well as from the $J$ and $H$
band images presented in McCabe et al. (2002). Their $K$ band
image is not used in this analysis as it was
inadvertently taken out of focus, preventing an accurate
estimate of the azimuthal surface brightness distribution of
the ring at that wavelength. The quantities we consider here
are the integrated disk/star ($R_\lambda$) and front/back
($FB_\lambda$) flux ratios of the ring, as well as the
distance between the center of mass of the system and the peak
of the scattered light flux ($D_{peak}$) in the ring's radial
profiles around position angle 7\degr, i.e., along its
semi-minor axis. We have also measured these same quantities
from the $HST$/WFPC2 images obtained by Krist et al. (2002)
with the F814W filter, a close match to the usual $I$
filter. Since this image has a very similar signal-to-noise as
the $J$ band image from McCabe et al. (2003), we assigned the
same uncertainties to both wavelengths. We now have an
homogeneous set of quantities that characterize the disk
surface brightness and that can be compared to each other.

We first consider the apparent geometry of the ring. The
ellipse that best fits the $L'$ image of the ring is in
excellent agreement with that fit by McCabe et al. (2002) to
the near-infrared images of the ring. Assuming that is has an
intrinsically circular shape, the observed aspect ratio of the
ring at $L'$ implies an inclination of about
40\degr$\pm$1\degr, somewhat larger than that derived from the
millimeter thermal emission (37$\pm$1\degr, Guilloteau et
al. 1999). As discussed in McCabe et al. (2002), this
difference is the result of the geometrical {\it and} optical
thickness of the ring. This implies that the ring is still
optically thick to the illuminating starlight at 3.8\mic.

\begin{figure}
\plotone{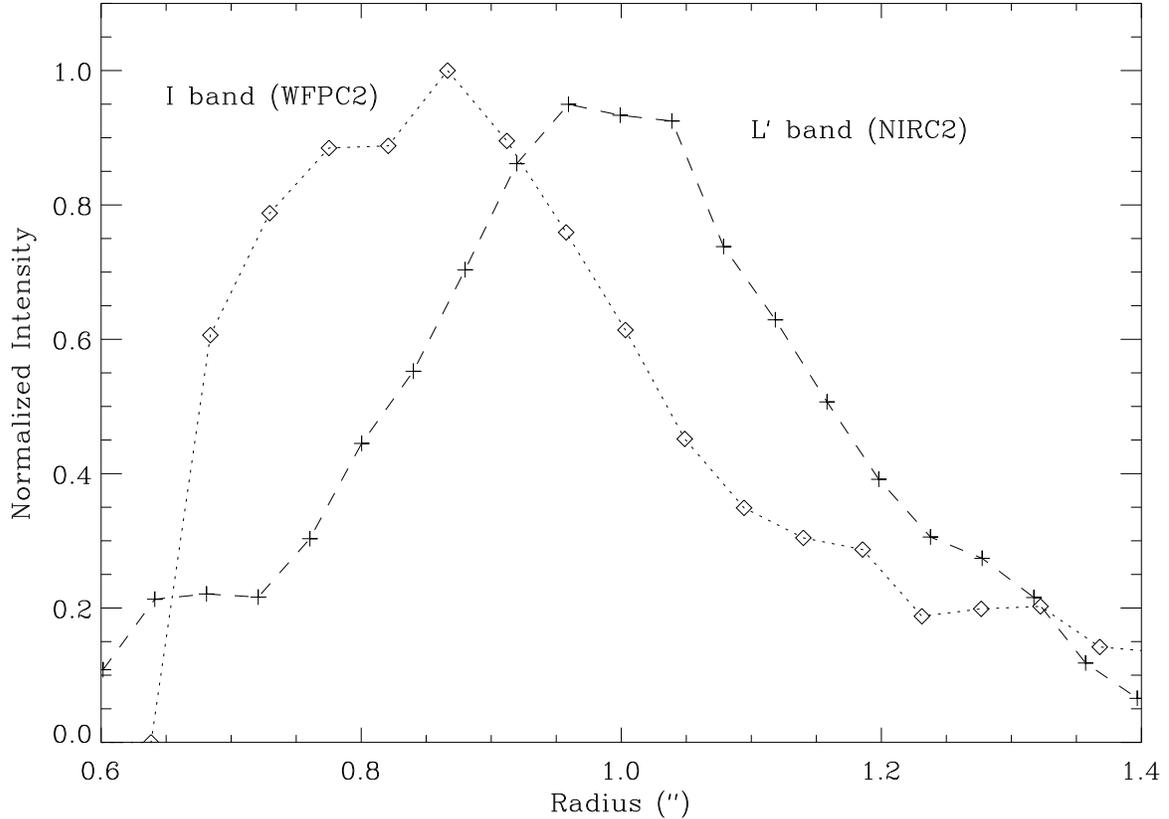}
\caption{Observed radial intensity profiles in a
10\degr-sector centered on the ring's northern semi-minor axis
(around position angle 5\degr). Radii are measured from the
center of mass of the system in both cases; the central
0\farcs6 of both profiles have been masked out as they are
dominated by PSF subtraction residuals. The diamonds and
dashed line represent the profile observed in the WFPC2 $I$
band image whereas the crosses and dashed line show the profile
from our $L'$ image of the ring. Both profiles are sampled with the
corresponding instrument's pixel scale (0\farcs0456 for WFPC2
and 0\farcs009942 for NIRC2).}
\label{fig:profileradial}
\end{figure}

Although the global shape of the ring is consistent with
previous images, a subtle though significant
wavelength-dependent effect can be noted in our image: the
distance between the center of mass of the system and the
northern edge of the disk along the semi-minor axis of the
system increases toward longer wavelength. This distance was
measured to be 0\farcs88 in the near-infrared by McCabe et
al. (2002). In the $I$ band image of Krist et al. (2002), we
measure this distance to be 0\farcs87, consistent with the
value measured in the near-infrared. On the other hand, the
equivalent distance in our $L'$ image\footnote{The NIRC2
detector has a well-characterized distortion pattern that
introduces a $\lesssim1$\,pixel, or 0\farcs01, error in a
$\sim$1\arcsec-distance measurement in any image of the ring,
which is negligible compared to the 0\farcs09 shift measured
here. Furthermore, since our final image is an average of many
images taken at 4 different locations on the detector and over
a $\sim180$\degr-range of orientations, we assume that the
distortion effect is actually even smaller.} is
0\farcs97. This shift, significant at the 5\,$\sigma$
confidence level (cf Table\,2), is clearly illustrated in
Figure\,\ref{fig:profileradial}. If the ring was geometrically
thin, scattered light would be detected at a distance
$R_{in}\cos i$ from the center of mass which can be inferred
from the dust millimiter thermal emission; both distances
measured in the scattered light images are significantly
smaller than this value. This offset has already been noted in
the past and convincingly interpreted as evidence that the
ring is optically thick up to high elevations above its
midplane (Guilloteau et al. 1999; McCabe et al. 2002). The
amplitude of the offset is proportional to the actual height
up to which the ring is optically thick. Our new result,
namely that this offset decreases toward longer wavelengths,
indicates that the dust opacity in the ring decreases enough
to bring the dominant scattering layer significantly closer to
the ring midplane. With $R_{in}=190$\,AU (see
\S\,\ref{subsec:geom}) and $i=37$\degr\ (Guilloteau et
al. 1999), we find that scattering of the
$\lambda\lesssim2\,\mu$m starlight toward the observer
predominantly occurs about 50\,AU above the ring midplane,
while it is only 25\,AU for the $L'$ photons. Note that the
offsets measured in the $I$, $J$ and $H$ images are consistent
with each other at the $\lesssim2\,\sigma$ confidence level,
indicating that all three wavelengths roughly probe the same
layer of the ring.

Besides the geometrical appearance of the ring, we also
searched for photometric changes with wavelength. The total
disk/star flux ratio displays significant variations from the
visible to the thermal infrared. The ring color excess indices
(with respect to the stars) are $\Delta
(I-J)=0.08\pm0.05$\,mag, $\Delta (J-H)=0.10\pm0.03$\,mag and
$\Delta (J-L)=-0.40\pm0.10$\,mag. Overall, we find that the
ring tends to become comparatively redder than the central
stars toward shorter wavelength, a trend that continues into
the visible ($\Delta (V-I)\sim0.8$\,mag, Krist et
al. 2002). The distribution of scattered light around the
stars, on the other hand, is almost independent of wavelength.
The position angle of the ring's brightest point at $L'$ is
fully consistent with that of shorter wavelength
images. Similarly, the undetected portion of the ring in our
image has consistently been found as the faintest in previous
high contrast images of the \ggt\ system. More generally, the
azimuthal intensity profile of the ring at $L'$ is remarkably
similar to those observed at shorter wavelengths, as
illustrated in Figure\,\ref{fig:profilepa}. The largest
chromatic effect we identify is the slight increase of the
peak-to-peak front/back flux ratio, which is typically 3--4 at
$I$, $J$ and $H$ (averaged over 5\degr- to
10\degr-sectors). Still, this effect is marginal.

Overall, the \ggt\ ring scattered light images appear to be
only weakly dependent on wavelength, except for the distance
between the center of mass of the system and the closest edge
of the ring, which is larger at 3.8\mic\ than at all other
wavelengths. As our Monte Carlo modeling shows
(\S\,\ref{sec:models}), this absence of strong chromatic
variations in the ring proves that grains larger than 1\mic\
are present, therefore that grain growth has occurred in this
environment (\S\,\ref{sec:discus}).

%________________________________________________________________

\section{Scattering models}
\label{sec:models}

Our goal is to devise a comprehensive model that reproduces
all scattered light images of the \ggt\ ring, namely the $I$ band
$HST$/WFPC2 image from Krist et al. (2002), the $J$ and $H$ $HST$/NICMOS
we have already obtained (McCabe et al. 2002) and the new $L'$
image presented here. Attempting to reproduce simultaneously visible,
near- and thermal infrared images of the ring will provide
stringent constraints on the dust grain properties, and
particularly their size distribution.

For this analysis, we have conducted a number of numerical
simulations that calculate the multiple scattering of photons
emitted from the central stars off the ring. We used the same
Monte Carlo code as in McCabe et al. (2002), and only present
here its most important input parameters and assumptions
(\S\,\ref{subsec:simu}). We then summarize the generic trends
revealed by these models as a function of wavelength and dust
grain size and identify a range of models that reasonably
reproduce each available image of the \ggt\ disk
(\S\,\ref{subsec:trends}).

% - - - - - - - - - - - - - - - - - - - - - - - - - - - - - - - - 

\subsection{Framework of the Monte Carlo simulations} 
\label{subsec:simu}

Our numerical model (M\'enard 1989; Duch\^ene 2000) randomly
propagates light packets (``photons'') in all directions from
the central object, which we treat for simplicity as a single
star having the integrated stellar light
properties\footnote{Krist et al. (2002) concluded that
including explicitly two stars in the simulations has only a
very marginal effect on the resulting synthetic images of the
\ggt\ ring.}. The circumstellar material density is tabulated
in a cylindrical coordinate system and we assume that the
system is azimuthally symmetric. Photons that reach the ring
are scattered after a random optical depth taken from an
$e^{-\tau}$ probability distribution. The size of the dust
grain that the photon scatters off is randomly selected in the
scattering cross section-weighted size
distribution\footnote{While more than 95\,\% of dust grains
have radii $a\leq0.1$\mic\ in all of our models, their
scattering cross section is so small that they do not account
for any significant scattered flux: between 95\,\% and 99\,\%
of all scattering events occur on grains larger than
0.1\mic. It is therefore not advisable to use a simple
``effective'' grain size, as absorption and scattering are
dominated by different dust grains.}, and the associated
Stokes transfer matrix is calculated through Mie theory. The
grains are assumed to be spherical or, if elongated, randomly
oriented so that their average cross-section is isotropic. At
each scattering event, photons lose a fraction (on the order
of the dust albedo) of their incoming intensity and photons
that have experienced more than 30 scatterings are terminated
as they carry a negligible intensity. We ran 4 million photons
per simulation and created synthetic intensity maps by
collecting all photons that leave the ring with an inclination
from the line of sight in the range 32--37\degr, similar to
the system's inclination (37\degr). Considering the 37--41\degr\
inclination range instead leads to similar conclusions; this
choice is therefore not critical for our analysis. The size of
the output images is about 4\farcs3 on a side, sampled at
0\farcs042/pixel.

To focus our analysis on the dust grain properties, we adopted
the same ring geometry for all models. We initially assumed
the ring geometry estimated from the (optically thin)
millimeter thermal emission by Guilloteau et al. (1999). The
relevant model parameters here are: the inner and outer radii
(180\,AU and 260\,AU), the total dust mass
(0.0013\,$M_\odot$), the vertical scale height (32\,AU at the
inner edge), the flaring index ($H(r) \propto r^{1.05}$) and
the surface density distribution ($\Sigma(r) \propto
r^{-1.7}$). Under the assumptions of vertical hydrostatic
equilibrium and isothermality, we tabulated the density
distribution of dust surrounding the binary in a
300\,AU-radius cylinder. We neglect the presence of the outer
disk seen in CO emission by Guilloteau et al. (1999) as the
ring itself is optically thick at all wavelengths considered
here, so that virtually no photons scatter off the outer
disk. We let the density distribution extend inside of the
ring inner radius with a gaussian function attenuation factor
described by a $1/e$ width of 7\,AU. The amount of material
located inside of 180\,AU therefore represents a small
fraction ($\sim5$\,\%) of the total disk mass. Because this
geometry does not provide a perfect match to the images of the
disk (see \S\,\ref{subsec:geom}), we increased the inner
radius to 190\,AU and reduced the $1/e$ width of the inner
radius cutoff to 2\,AU. These are the values used in all the
models discussed quantitatively in the following.

We assumed the composite, porous dust grain properties from
Mathis \& Whiffen (1989, their model A), resulting in a
0.5\,g/cm$^3$ average grain density. The numerical models are
monochromatic with the central wavelength of the filters (0.8,
1.0, 1.6 and 3.8\mic\ respectively for the $I$, $J$, $H$ and
$L'$ filter) used to match the various observations of the
ring. Chromatic changes in the scattering properties within a
given bandpass is a second order effect that we neglect
here. The dust grain size distribution is assumed to be a
single power law, $N(a) \propto a^{-3.7}$, as derived by
Mathis \& Whiffen for interstellar dust grains, but with
different truncation values. The smallest grain size is taken
to be $a_{min}=0.03$\mic\ (corresponding to a size parameter
of $0.05 \lesssim x \lesssim0.25$) in all models since even
smaller grains do not contribute to scattering at visible or
near-infrared wavelengths. With respect to the value quoted by
Mathis \& Whiffen ($a_{min}=0.005$\mic), this only has an
impact on the average dust opacity but the latter is not
modified by more than 15\,\%. Since we are interested in
constraining the dust grain size distribution, its upper
cutoff, $a_{max}$, was the only parameter that we let vary in
our models\footnote{The slope of the grain size distribution
is another important parameter describing the dust
population. Varying this slope could result in slightly
different values for the best fitting $a_{max}$, but the
overall behavior describe below would remain the same.}. We
used values of $a_{max}$ ranging from 0.25\mic\ to
3.0\mic. For each value of $a_{max}$, the average mass of the
grains over the entire size distribution is calculated and the
total number of grains is then set so that the total dust mass
in the ring remains the same in all simulations.

A complete list of the Monte Carlo simulations we ran is
presented in Table\,3. Five models (3, 8, 12,
15 and 17) were calculated at all four wavelengths while 4 to
6 additional models were run at each wavelength to ensure a
fine sampling around the best fit $a_{max}$. Overall, 42
different simulations were run. Each synthetic image has been
convolved with an appropriate Gaussian kernel so that it has a
similar resolution as the observed one.

% - - - - - - - - - - - - - - - - - - - - - - - - - - - - - - - - 

\subsection{Results and general trends from the simulations}
\label{subsec:trends}

The intensity maps for all models that were calculated at all
wavelengths are presented in Figure\,\ref{fig:models}. We can
readily see that no single model can reproduce simultaneously
all the images of the ring, since longer wavelength images
seem to require larger values of $a_{max}$ than the $I$ and
$J$ images. Before comparing in more detail the models to the
observations, we first describe here some generic trends that
are nicely illustrated in Figure\,\ref{fig:models}.

The natural dependency of scattering on the ratio of dust
grain size to wavelength is clearly illustrated by our
numerical simulations. Two models with almost equal
$a_{max}/\lambda$ ratio (for instance model 3 at $J$ band and
model 8 at $H$ band) have very similar synthetic
images. Generally speaking, the shorter the wavelength and/or
the larger $a_{max}$, the more forward-throwing the
scattering. Since the ring is not seen exactly pole-on but
tilted by about 37\degr, forward-throwing scattering results
in more flux on the front (Northern) edge of the ring, as is
observed in most of our models. The asymmetry parameter, $g$,
is the quantity that measures the preference for
forward-throwing scattering in Mie theory. Isotropic
scattering has $g=0$ while purely forward-throwing dust grains
are characterized by $g=1$. For each of our models, we define
a median asymmetry parameter, $\overline{g_\lambda}$ in the
following way: from the scattering probability distribution,
we determine what the ``median scatterer'' size ($a_{med}$)
is, defined so that exactly 50\,\% of all scattering events
occur on larger grains. The asymmetry parameter corresponding
to this grain size is taken to be $\overline{g_\lambda}$; the
value of this parameter for all models are listed in
Table\,4. As an illustration of isotropic scattering, one can
see in Figure\,\ref{fig:models} that the $L'$ images of models
that only contains small grains ($a_{max} \lesssim 0.5$\mic;
those models have $\overline{g_{L'}}<0.1$) have their peak
surface brightness on the back (southern) side of the ring,
which presents a larger solid angle than the front edge of the
ring, as opposed to most other models and to all actual images
of the ring.

Our numerical simulations predict significantly redder colors
for the ring than for the illuminating stars for almost all
models, up to $\sim0.5$\,mag color excess despite neutral or
slightly blue albedo values. We already pointed out this
effect in McCabe et al. (2002) and suggested that this was a
consequence of the high and strongly wavelength-dependent
values of $\overline{g_\lambda}$ for the Mathis \& Whiffen
(1989) dust model. Here, we find that the models with the
larger $a_{max}$ systematically have larger color indices, in
agreement with this explanation since they have larger
asymmetry parameters. 

For each wavelength, we can determine which model best matches
the observations. We are particularly interested in analyzing
the scattering phase function of the dust grains in the ring,
as the latter is strongly dependent on the size of the largest
grains in the ring. There are several possible quantitative
approaches to comparing the synthetic maps produced by our
Monte Carlo simulations and the observed images of the
ring. Both the disk/star flux ratio ($R_\lambda$) and
integrated front/back ratio ($FB_\lambda$) are very well
defined in the simulations. Therefore, we computed both of
these flux ratios for all of the simulations using the same
elliptical aperture as defined in \S\,\ref{sec:results}. The
results are tabulated in Table\,4; relative
uncertainties are less than 2\,\% as measured from the
standard deviation of measurements for two models that were
run three times.

While both $R_\lambda$ and $FB_\lambda$ depend on the dust
scattering phase function, they also are strongly sensitive to
the assumed disk geometry. In particular, they depend on the
solid angle of the ring as seen from the central star and,
because the ring is both optically and geometrically thick, the
effect is not the same on the front and back side of the ring.
Since we would like to focus on the scattering properties of
dust grains, it is not advisable to use quantities that can be
strongly affected by uncertain assumptions on the disk
geometry. Integrated fluxes are therefore not the best
quantities to use in comparing the models to the
observations. The peak azimuthal brightness variations offers
a better handle on the grains properties, as it basically
traces the single-scattering intensity off an optically thick
wall, i.e., essentially the dust scattering phase function. It
is much less sensitive to geometric assumptions than
integrated flux ratios as scattering angles ranging from
$\sim50$\degr\ (on the norther side of the ring) to
$\sim120$\degr\ (on its southern part) are probed. Fitting the
entire azimuthal surface density profile is therefore more
appropriate than integrated fluxes or than peak-to-peak flux
ratios, which can be corrupted by local perturbations in the
structure of the ring.

\begin{figure*}
\plotone{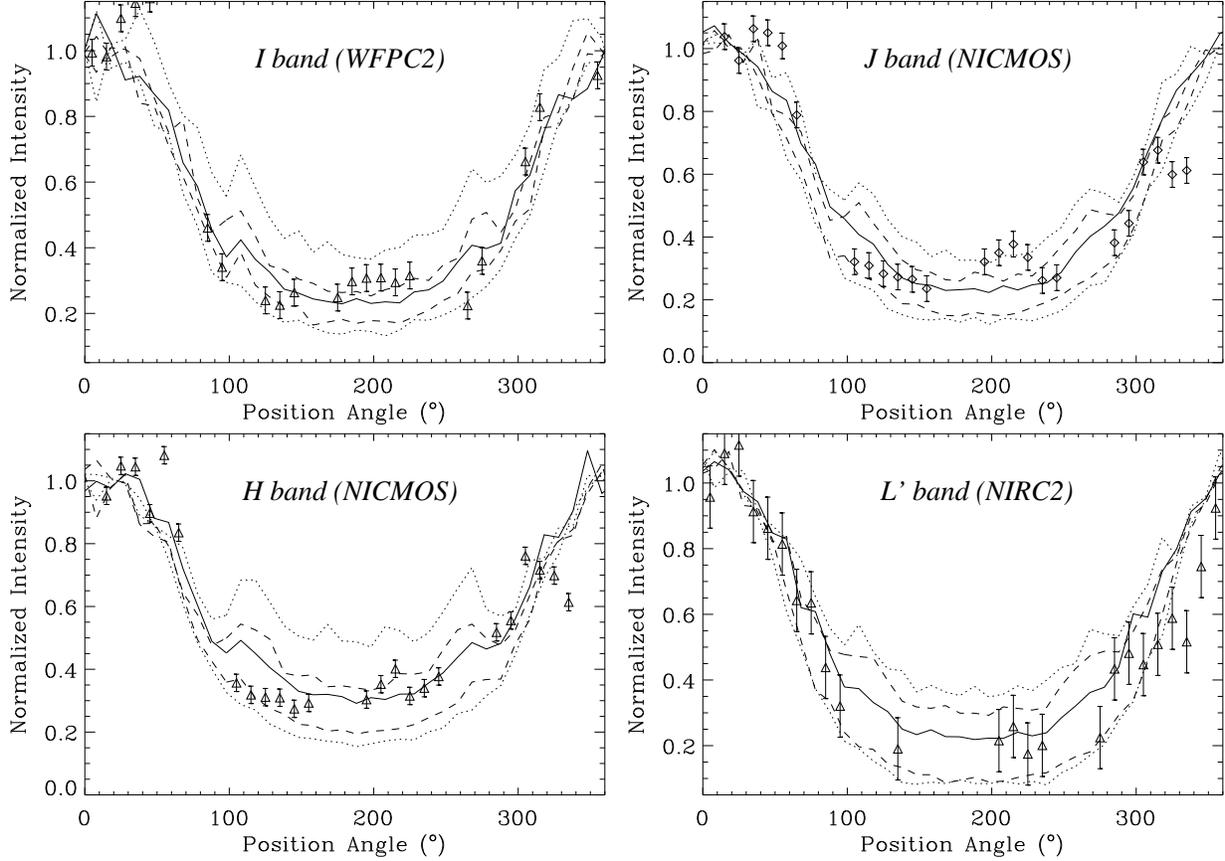}
\caption{Azimuthal intensity profiles of the \ggt\
circumbinary disk. Each profile has been normalized to the
average flux in the position angle range 345\degr--25\degr,
which is centered on the disk semi-minor axis. The solid
curves indicates our best matching Monte Carlo simulation
while dashed and dotted lines indicate the range of acceptable
and the closest unacceptable models, respectively (see
\S\,\ref{subsec:trends}).}
\label{fig:profilepa}
\end{figure*}

We therefore adopted the following method, which is based on
the azimuthal intensity variations and is illustrated in
Figure\,\ref{fig:profilepa}, to determine a range of
acceptable models. For each wavelength, we first plotted the
observed intensity variations, normalized to the average flux
in the position angle range (centered on the disk's semi-minor
axis) 345\degr--25\degr. Then all models calculated at that
wavelength were overplotted and normalized in the same way. A
modified $\chi^2$ value was calculated for all models, based
on the uncertainties in the observed
images\footnote{Uncertainties were first calculated for each
10\degr\ sector as the standard deviation of the measurements
from 3 or 4 sub-images, then averaged over all sectors in a
given image of the ring and uniformly assigned to all
measurements.} and the typical uncertainty of the numerical
simulations, estimated from the standard deviation of the
profiles of three independent simulations of the same
model. The model that results in the smallest $\chi^2$ is the
best match to the observations. We consider models with
$\chi^2$ values that are not more than 30\,\% larger than the
best models to be acceptable models. In the absence of a
proper characterization of the uncertainties' statistical
properties (for both the models and the observations), we did
not adopt this threshold based on a robust statistical
analysis but we consider that it defines a reasonably
conservative range of acceptable models, probably equivalent
to a $\sim3\,\sigma$ confidence level, as illustrated in
Figure\,\ref{fig:profilepa}. Models out of that range usually
result in very poor fit of the azimuthal intensity variations.

In order of increasing wavelength, the $I$ band image of the
ring is best reproduced by models 2, 3 and 5 (0.275\mic$\leq
a_{max}\leq$0.35\mic), the $J$ band image by models 4--7
(0.325\mic$\leq a_{max}\leq$0.45\mic), the $H$ band image by
models 7--10 (0.45\mic$\leq a_{max}\leq$0.6\mic) and the $L'$
band image by models 14--19 (1.125\mic$\leq
a_{max}\leq$2.5\mic). As illustrated in
Figure\,\ref{fig:models}, increasingly large dust grains are
required to reproduce the images at longer wavelengths. The
best fit models systematically underestimate $FB_\lambda$ and 
overestimate $R_\lambda$. Using
either of these two flux ratios would therefore yield a different
set of best-fitting dust model. However, the general behavior
of increasing $a_{max}$ values being required to reproduce
longer wavelength images would still hold. As explained above,
we believe that our method is the best to constrain the scattering
phase function of dust grains, hence their size, because it is
less dependent on the assumed geometry of the ring.

For the models calculated at all wavelengths, we find that the
distance between the front edge of the ring and the center of
mass of the binary systematically increases with wavelength,
as a consequence of the decreasing opacity. For all models,
the ratio of opacities is on order of $\kappa_{L'}/
\kappa_I \approx10$, so that scattering at 3.8\mic\ probes
much deeper into the ring. A given $\tau_\lambda$ surface is
therefore closer to the disk midplane at longer wavelength
and, given the inclination of the ring, it appears closer to
the projected inner radius of the ring, i.e. further away from
the center of mass of the system. This predicted behavior is
also present in the data (see \S\,\ref{subsec:ring_all}) and
the simulations predict an amplitude for this effect of
0\farcs05--0\farcs08, in rough agreement with our geometrical
model of the ring.

%________________________________________________________________

\section{Discussion}
\label{sec:discus}

In light of our multi-wavelength analysis of the \ggt\ ring,
both the geometry of the ring and the properties of the dust
population in the ring can be studied in much more detail than in
previous work. We first show that the slightly
revised disk geometry adopted here better matches the observed
images (\S\,\ref{subsec:geom}) and then argue that grains
larger than those of the interstellar medium are present in
the ring, in a stratified configuration
(\S\,\ref{subsec:dust}).

% - - - - - - - - - - - - - - - - - - - - - - - - - - - - - - - - 

\subsection{Revisiting the geometry of the system}
\label{subsec:geom}

Most scattered light models of the \ggt\ ring are based on the
geometry derived from the optically thin millimeter thermal
emission by Guilloteau et al. (1999), namely an axisymmetric,
low-flaring disk with very sharp inner and outer radii. Here
we review some evidence that this geometry might require minor
revisions to reproduce the images of the ring.

First, all scattered light images of the system have shown
that the ring is not completely smooth and that there are some
departures from simple symmetries, such as the ``gap'' clearly
shown by Krist et al. (2002) on the western side of the ring
or the displacement between the ring's brightest point and it
semi-minor axis. These could be the result of local density
perturbations, such as a gravitational wave at the inner edge
if the disk, or of shadowing effects induced by obscuring
material located around the stars themselves. Although these
perturbations are strong enough to create clear asymmetries in
the scattered light images, the absence of azimuthal
variations in the optically thin thermal emission of the ring
suggest that departures from axisymmetry must be of relatively
small amplitude. In our model, we have kept the axisymmetry
assumption because it allows a dramatic decrease in the number
of photons needed to run our Monte Carlo simulations thereby
enabling a better sampling of the parameter space. Because we
use the complete azimuthal intensity variations (using
averages within 10\degr\ sectors) to compare models to the
observed images of the ring, we consider that small scale
intensity variations are unlikely to strongly affect our
conclusions regarding dust properties.

In our previous modeling effort of the \ggt\ ring, we had
used essentially the same numerical code and disk geometry,
except that we had enforced a strict vertical cutoff to the
ring at about 35\,AU above the disk midplane (McCabe et
al. 2002); no dust grains were present above such an
elevation. This was required to match the observed distance
between the center of mass of the system and the Northern edge
of the ring. Here, we have not used such a cutoff, so that a
decreasing amount of material is located even further high
above the midplane. There are two direct consequences to the
absence of vertical cutoff: i) the synthetic scattered light
images have the front edge of the ring too close from the
stars by about 0\farcs10, and ii) the integrated disk/star
flux ratio is too large by a factor of 2--3 with respect to
the observations, as opposed to the results of the models with
a cutoff. A vertical cutoff therefore appears to be necessary
to match the location of the front edge of the ring and the
total disk/star flux ratio. However, numerical simulations
that include it fail to reproduce the extent of the back side:
in all simulations with a cutoff, the back side of the ring
does not extend beyond $\sim1$\farcs2 of the center of mass
along the semi-minor axis (see
Figure\,\ref{fig:comparmodels}). Since both the visible and
near-infrared images clearly show that the back side actually
extends $\sim$1\farcs7 away from the center of mass, we reject
the hypothesis that such a vertical cutoff is actually present
in the \ggt\ ring.

In the ``no-cutoff'' model we argue for here, we believe that
the disagreement between the predicted and observed values of
$R_\lambda$ can be explained by improper albedo values, which
depend more on the dust composition than the scattering phase
function, and/or by an unrealistic treatment of
self-absorption toward the front edge of the ring, where
most of the ring flux comes from. For instance, the vertical
distribution of dust may not be gaussian, especially at the
ring's inner edge where a pile-up of material can be expected
due to the detailed physics of the gravitational resonance
induced by the orbital motion of the binary. For the time
being, we believe that this issue does not represent a major
shortcoming of our model of the ring. However, the observed
distance between the system's center of mass and the ring
front edge is significantly larger than the ``no-cutoff''
model predicts as pointed out above.  To reconcile the models
with the observations on this aspect, we have increased the
ring's inner radius from 180\,AU to 190\,AU and shrank the
$1/e$ width inner density drop-off from 7\,AU to 2\,AU. As
illustrated in Figure\,\ref{fig:comparmodels}, this results in
an excellent match between the observed and predicted
images of the ring. This revision to the ring's inner radius is only
marginal and is within 2$\,\sigma$ of the best-fitting values
derived by Guilloteau et al. (1999). Furthermore, our
numerical simulations show that the increased ring's inner
radius only results in a $\sim7\,\%$ loss in the predicted
scattered light flux from the ring, i.e., a negligible effect
for our analysis. We therefore adopt this new set of
parameters in our present analysis.

Another important aspect regarding the geometry of the system
is the possible presence of some obscuring material between
the binary system and the ring. The observed neutral/red
excess color indices of the ring with respect to the stars has
led Roddier et al. (1996), Wood et al. (1999) and Krist et
al. (2002) to argue that such material is present in order to
compensate the intrinsically blue color of
scattering. However, as discussed in \S\,\ref{subsec:trends},
scattering off the optically thick ring can result in red
colors depending on the assumed dust properties. Using our
best fit models for each image of the ring, which have been
determined independently of the ring's flux and color indices,
we predict color excesses with respect to the central sources
of $\Delta (I-J) = 0.06$\,mag, $\Delta (J-H) = -0.01$\,mag and
$\Delta (H-L') = -0.11$\,mag, in reasonable agreement with the
observed values to within 0.1--0.2\,mag. This suggests that
there is only little extinction between the stars and the
ring. The large $\Delta (V-I)=0.8$ \,mag color excess found by
Krist et al. (2002) supports a moderate extinction in the
plane of the ring. Their $V$-band image of the ring lacks the 
signal-to-noise to allow model fitting unfortunately, so we cannot
determine the best-fitting model at that wavelength. Still,
{\it if} the region of the ring probed by scattered light at
$V$ is identical to that probed at $I$ (or if those two
regions contain the same dust population), we predict a color
excess of $\Delta (V-I)=0.28$\,mag (see
Table\,4), which is not as red as the
observations of the ring. The remaining color excess, $\Delta
(V-I)\sim0.5$\,mag, corresponds to an in-the-plane extinction
of $A_V\approx1.0$\,mag assuming an $R_V=3.1$ extinction law
(Rieke \& Lebofski 1985). Overall, the observed colors of the
\ggt\ ring are consistent with a system's geometry in which
there is a moderate amount of reddening between the stars and
the ring. Note that the absence in all images of the ``dark
central lane'' on the back side of the ring, as predicted by
Wood et al. (1999) assuming coplanar optically thick
circumstellar disks, also requires that extinction in the
plane of the ring must be optically thin.

% - - - - - - - - - - - - - - - - - - - - - - - - - - - - - - - - 

\subsection{A stratified dust population}
\label{subsec:dust}

As pointed out in \S\,\ref{subsec:ring_all}, the azimuthal
distribution of scattered light is remarkably similar in all
images of the ring over a factor of 4 in wavelength, from $I$
to $L'$. This suggests that the scattering properties of the
dust grains, in particular their phase function, evolves only
marginally over this wavelength range. This is in sharp
contrast with all models of ISM grains, which have
$a_{max}\leq1$\mic\ and, therefore, are much more forward
throwing in the visible than in the near-infrared (Kim et
al. 1994; Zubko, Krelowski \& Wegner 1998). In the following,
we use the range of acceptable phase function asymmetry
parameters we found at each wavelength
(\S\,\ref{subsec:trends}) to constrain the dust grain size
distribution in the ring and we argue that it is likely that
the ring has a stratified structure, in which larger grains
are present closer to the disk midplane.

First of all, an important qualitative statement can be
readily made in the framework of our assumed dust model: since we find
no value of $a_{max}$ that matches 
all images of the ring simultaneously, we conclude
that the size distribution in the ring is not a simple power
law, and/or that the dust population in the ring is not
uniform. The latter explanation would be the correct one if
each wavelength was probing a different region of the disk
with its specific size distribution. While we found that the
$L'$ image of the ring probes deeper into the disk than
shorter wavelength images, we have also shown that all images
at 0.8\mic$\leq \lambda \leq$1.6\mic\ probe the same region of
the ring, located about 50\,AU above its midplane. However,
the acceptable models for reproducing the $H$ band image
disagree significantly with the $I$ and $J$ images of the
ring. Therefore, we can exclude that the surface layer of the
ring contains a single power law size distribution.

\begin{figure}
\plotone{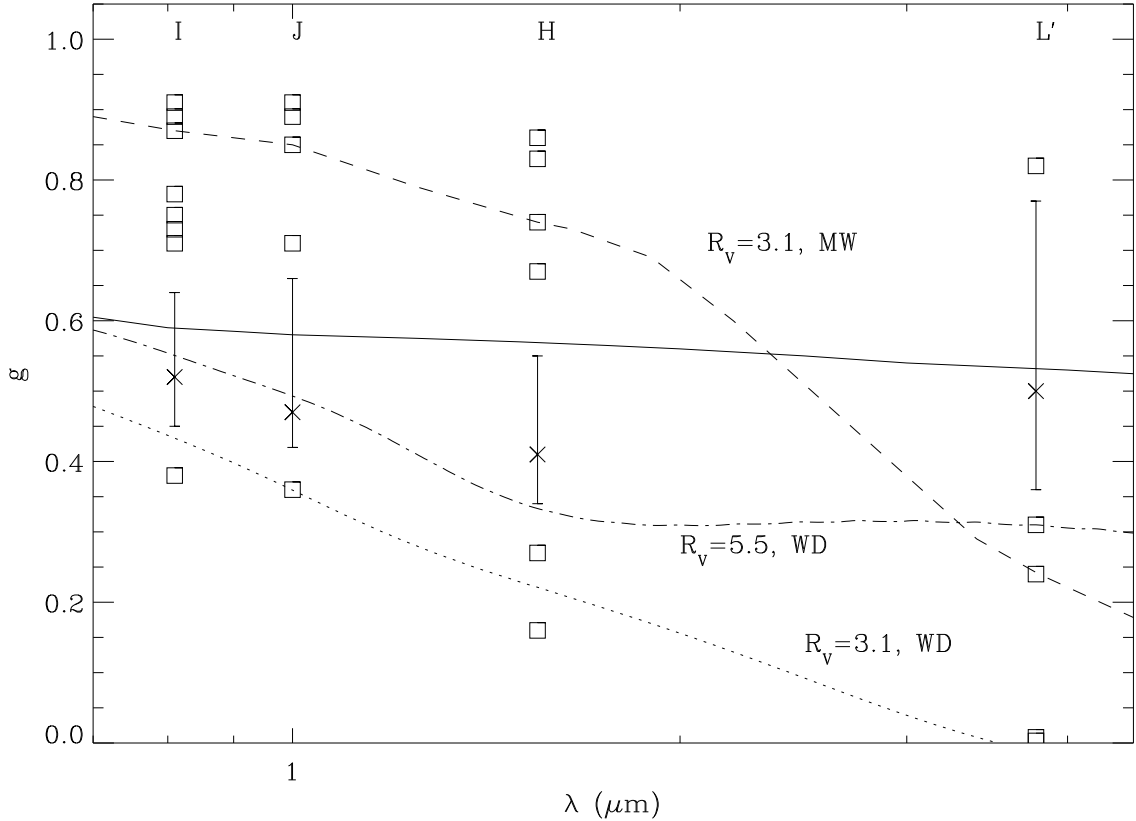}
\caption{Fitted values of the phase function asymmetry
parameter for the dust grain population in the \ggt\
ring. Crosses indicate the best matching models and the
acceptable ranges are indicated as solid segment; these
represent $\sim3\,\sigma$ confidence intervals. Open squares
are models that can confidently be excluded. The dashed line
is the predicted dependence of $g$ for the dust grain models
developed by Mathis \& Whiffen (1989) to match the
interstellar ($R_V=3.1$) extinction law. The dotted and
dot-dashed curves are the predicted behaviors of the dust
models developed by Weingartner \& Draine (2001) for
$R_V=3.1$ and $R_V=5.5$ extinction laws, respectively. The
solid curve is the dust model derived by Wood et al. (2002) to
match the spectral energy distribution of the edge-on
circumstellar disk system HH\,30. Note that the various
wavelengths analyzed here for \ggt\ do not necessarily probe a
single dust population.}
\label{fig:g}
\end{figure}

Before exploring the possibility of spatially-dependent grain
properties, we first attempt to find a refined dust model that
reproduces all the images of the ring under the assumption of
an homogeneous dust population throughout the ring. In
particular, we would like to find a dust model that matches
the entire $\overline{g_\lambda}$ curve derived for the \ggt\
ring. While our models of the ring, which assume azimuthal
symmetry and power law grain size distributions, may be too
simplistic, we believe that we are properly probing the
scattering phase function of the dust grains, so that our
derived values of $\overline{g_\lambda}$ are probably close
matches to that of the actual grain population\footnote{Krist
et al. (2002) derived $\overline{g_I}=0.65$ from their image
of the ring, which is just outside of the range we have
accepted here. While this is probably not significant, we note
that their best fit value is obtained by matching the maximum
peak-to-peak flux ratio between the front and back side of the
ring, whereas we consider the entire azimuthal intensity
variations. These slightly different methods probably explains
the small disagreement between the two best fit values of
$\overline{g_I}$. Note that our method is intrinsically less
likely to be affected by small scale intensity variations in
the ring.}. Figure\,\ref{fig:g} summarizes our results
regarding the dust grains in the \ggt\ ring, as well as some
frequently used ISM dust models (Mathis \& Whiffen 1989;
Weingartner \& Draine 2001). It is important to realize that,
for these ISM dust models, the scattering phase function is
calculated after a model is fit to the general ISM {\it
extinction/emission} law defined by $R_V=3.1$, whereas our
analysis of the \ggt\ ring dust population provides {\it
direct} estimates of $\overline{g_\lambda}$. In this context,
it is noteworthy that two different models reproducing the
same ISM extinction law can have asymmetry parameters that
differ so much. This results from the different assumptions
made on the dust composition and shape of the size
distribution but, as shown in Figure\,\ref{fig:g}, this raises
doubts about which model best matches the ISM dust
population. Independently of this unsolved ambiguity, the
observed shape of the $\overline{g_\lambda}$ curve for the
\ggt\ ring cannot be matched by any current ISM dust
model. Larger dust grains are required to match the value of
$\overline{g_{L'}}$ derived in this study, as evidence by the
1.125--2.5\mic\ upper limit for the size distribution we have
found assuming a simple power law for the size
distribution. Even larger grains may be present in the ring
provided that the size distribution has a more complex shape,
such as the $N(a)\propto a^{-p} e^{-(a/a_c)}$ form sometimes
used for ISM models (e.g., Kim et al. 1994; Clayton et
al. 2003).

This need for a significant population of grains larger than
1\mic\ is reminiscent of the dust models developed to
reproduce the extinction curves measured through molecular
clouds, which are likely to contain the precursor material of
future circumstellar/binary disks. As illustrated in
Figure\,\ref{fig:g}, the model derived by Weingartner \&
Draine (2001) to reproduce an $R_V=5.5$ extinction law
provides a good match to the asymmetry parameters at $I$ and
$J$ and is only slightly off the range of acceptable models at
$H$. However, this model predicts too low a value for
$\overline{g_{L'}}$, even though it contains grains as large
as 10\mic. In other words, the grain size distribution
inferred for molecular cloud does not contain {\it enough}
grains larger than 1\mic\ to reproduce our $L'$ image of the
\ggt\ ring, suggesting that further growth of the dust grains
(or a relative depletion of the smaller particles) has occurred
in the ring.

The only dust model we are aware of that resembles the
observed wavelength dependence of $\overline{g_\lambda}$ is
the model that Wood et al. (2002) constructed to reproduce the
entire spectral energy distribution of HH\,30, a young stellar
object that is surrounded by an opaque, edge-on circumstellar
disk. This model does not provide a perfect match however, as
it systematically overestimates the asymmetry parameters at
wavelengths $\lambda\lesssim1.6$\mic\
(Figure\,\ref{fig:g}). Still, this suggests that a model in
which the dust population of the ring is spatially uniform and
has a size distribution similar to that found by Wood et
al. for another equally young object, i.e., extending up to
$\sim1$\,mm, could reproduce the various images of the \ggt\
ring.

An independent empirical test can be performed regarding the
dust population of the \ggt\ ring based on the $HST$/NICMOS
1\mic\ polarization map obtained by Silber et al. (2000),
which probes the surface layer of the ring. The dust grain
size distribution not only dictates the scattering phase
function, it also determines the maximum linear polarization
rate for a single 90\degr-scattering event. Recently, Whitney
et al. (2003) have calculated the maximum
polarization rates corresponding to ISM and molecular cloud
dust grains, as well as to a dust model containing somewhat
larger grains that has been devised to reproduce the
near-infrared scattered light images of HH\,30 (Cotera et
al. 2001) and, finally, to the dust model proposed by Wood et
al. (2002). At 1\mic, they find a maximum polarization rate of
50-55\,\% for both ISM and molecular cloud-like dust models
but only $\sim40\,\%$ for the latter two models. The lower
polarization rate in the models containing larger dust grains
is a consequence of the larger polarizing efficiency of small
dust grains in the Rayleigh regime. The polarization map of
Silber et al. (2000) peaks at $\sim50\,\%$ for a scattering
angle of $\sim100$\degr. This must be considered as a lower
limit to the actual single-scattering maximum rate in
particular because of the diluting effect of multiple
scattering on the linear polarization. Therefore, only ISM-
and molecular cloud-like dust models can account for such high
polarization rates as observed in the ring, ruling out the
possibility of much larger grains in its surface layer.

To summarize, we find that a dust model similar to those
derived for molecular cloud environments can simultaneously
reproduce the images of the ring at all wavelengths shorter
than 1.6\mic\ as well as the 1\mic\ polarization map of the
\ggt\ ring. However, the new $L'$ image of the ring presented
here requires much larger dust grains, supporting a model that
is on the other hand unable to reproduce all of the shorter
wavelength constraints. More explicitly, it appears that the
$\sim50\,\%$ linear polarization rate observed at 1\mic\ is
incompatible with the highly forward-throwing scattering
observed at 3.8\mic. Combined with the fact that the $L'$
scattered light image probes a different layer of the ring
than the shorter wavelength datasets (\S\,\ref{subsec:geom}),
this implies that the dust grain properties are not uniform
throughout the ring.  Namely, the surface layer of the ring,
located $\sim50$\,AU above its midplane, contains a dust
population that is very similar to that observed in dense
molecular clouds whereas the layer of the ring located twice
as close to the midplane, which is probed by our new thermal
infrared image, contains significantly larger grains. Exactly
how large the dust grains are depends on the exact shape of
the size distribution and on the dust composition. Indeed, a
single power-law distribution with $a_{max}\sim1.5$\mic\ and a
Mathis \& Whiffen (1989) composition leads to essentially the
same value of $\overline{g_{L'}}$ as the amorphous carbon and
silicate model of Wood et al. (2002) that extends up to 1\,mm
in radius. Although we cannot yet fully determine the dust
grain properties in the deeper layers of the ring, we can tell
that they contains dust grains larger than $\sim$1\mic\ in size 
in non-negligible amounts, which is not true for the surface of
the ring.

Interestingly, the shallow opacity law for the ring
($\kappa_\nu \propto \nu^\beta$, with $\beta = 1.15\pm0.10$)
found by Guilloteau et al. (1999) is similar to that predicted
by the dust model of Wood et al. (2002), another argument
favoring this dust model for the bulk of the dust component of
the ring. This is indeed one of the first times that a dust
model with very large grains ($a_{max}\sim1$\,mm), such as
frequently supported by thermal millimeter emission, is
consistent with a scattered light image. In other well-studied
T\,Tauri disks, like those of HH\,30 or TW\,Hya, only the long
wavelength thermal emission requires such large grains to be
present, whereas all scattered light images are well
represented by dust grains that are not significantly larger
than those found in molecular clouds (e.g., Burrows et
al. 1996; Cotera et al. 2001; Weinberger et al. 2001).

The only other disk for which scattered light images have
called for much larger dust grains than the ISM is HK\,Tau\,B,
whose scattered light disk as seen in the visible is
well reproduced with ISM-like dust grains (Stapelfeldt et
al. 1998) while a recent mid-infrared scattered light has led
to the conclusion that it also contains much larger grains
(McCabe et al. 2003). While the analysis of the thermal
emission from this object is still under debate (D'Alessio et
al. 2001; Duch\^ene et al. 2003), we note that the dust model
developed by Wood et al. (2002) is fully consistent with the highly
forward-throwing scattering properties observed at 12\mic\ by
McCabe et al. (2003). The fact that only long wavelength
($\lambda>3$\mic) scattered light images of protoplanetary
disks require the presence of multi-micron dust grains is in
part a consequence of the wavelength-dependence of scattering. 
However, it can also be interpreted as evidence
that the surface layers of protoplanetary disks usually
contain ISM and/or molecular cloud-like dust populations
whereas the deeper regions of these disks, probed by
millimeter thermal emission and now also by thermal- and
mid-infrared scattering, contain much larger grains. In this
picture, we predict that long wavelength scattered light
images of other protoplanetary disks such as HH\,30 or TW\,Hya
will reveal highly forward-throwing, hence much larger, dust
grains.

Stratified disk models have already been studied in the
context of protoplanetary disks (e.g., Chiang et al. 2001;
Lachaume et al. 2003) and have been proposed for several
systems (Weinberger et al. 2002; Duch\^ene et al. 2003). The
physical cause for this stratification of the ring could be i)
preferred dust coagulation in the disk midplane, where the
density and therefore the grain-grain interaction probability
are highest, and/or ii) gravitational settling of the largest,
hence most massive, grains which cannot be prevented from
falling toward the midplane by turbulence or gas heating. It
is not possible for now to distinguish between these two
options, except for the fact that we have shown that the
surface layers of the \ggt\ ring contains almost no grains
larger than those found in molecular cloud, i.e., they appear
to constitute an unevolved dust population. If larger grains
had grown there, it may be difficult to explain why only those
grains that have formed in the process have sunk unless
turbulence in the disk is strong enough to maintain the small
grains in suspension high above the midplane. On the other
hand, our best model of the ring implies that the volume
density of grains is only 5 times larger at an elevation of
25\,AU with respect to the (surface) 50\,AU layer. It is
unclear whether such a modest change in density is enough to
trigger significant grain growth through grain-grain
interaction.

From the similarity between the dust properties contained in the
ring's surface layer and those of a typical molecular cloud, one may
conclude that only marginal grain growth has occurred in the
pre-collapse dense core that led to the formation of the
binary system. In this context, the dust grains probed by the
shortest wavelength scattered light images can be considered
as pristine, while the larger dust grains required to explain
our observed $L'$ image of the ring, as well as its millimeter
thermal emission properties, imply that grain growth has
occurred deeper in the ring, in its highest density regions,
after it was initially formed. While this may not seem
surprising, it is noticeable that, as opposed to most young
disks in which grain growth has been suggested in the past, we
extend this conclusion to a much wider, circumbinary
structure, whose dynamical timescale is on order of 2000\,yrs
instead of a few decades. One could therefore have expected
grain growth to be a much slower phenomenon in this ring. For
instance, the dust evolution model of Suttner \& Yorke (2001),
which can lead to very large grains in $\sim10^4$\,yrs within
30\,AU of the central star, predicts almost no change of the
dust size distribution at a radius of 200\,AU over the same
timescale. The dynamical timescale is about 20 times longer at
this radius, so a similar growth could require a few
$10^5$\,yrs, only a few times younger than the \ggt\ system itself.
Still, our thermal infrared image of the system suggests
that significant grain growth has already occurred in the ring.

%________________________________________________________________

\section{Conclusion}
\label{sec:concl}

Using the AO system on the W. M. Keck\,II 10\,m-telescope, we
have obtained the first 3.8\mic\ scattered light image of the
circumbinary dusty torus around the young binary system
\ggt. This is the longest wavelength at which the ring has
been imaged in scattered light so far. The ring surface
brightness variations appear to be remarkably similar to those
observed in the visible and near-infrared in previous $HST$
studies of the system, suggesting that the scattering
properties of dust grains in the ring are barely dependent on
wavelength, a distinctly different behavior when compared to
ISM dust models. We also find that the scattered light from
the front edge of the ring, which traces the upper optically
thick layers above the ring's midplane, is projected further
away from the center of mass of the system at $L'$ than in
shorter wavelength images. This is consequence of the reduced
dust opacity at 3.8\mic\ and it allows us to probe for the
first time the dust scattering properties only 25\,AU above
the midplane, instead of 50\,AU above it in previous studies.

Using a Monte Carlo multiple scattering numerical code, we
have conducted a multi-wavelength modeling analysis of the
scattered light images of the ring in an attempt to study in
more detail both the geometry of the ring and the properties
of the dust grains it contains. We find that all images are
best reproduced with a slightly increased inner radius
(190\,AU instead of 180\,AU) and steeper density fall-off
inside of this radius. Our best fit models, although they
systematically predict $\sim3$ times too much scattered flux
from the ring, result in excess color indices with respect to
the central stars from $I$ to $L'$ that are consistent with
the observations within 0.1--0.2\,mag typically. The predicted
$V-I$ color of the ring falls short of the observed one,
however, suggesting that there is a limited amount of
reddening ($A_V \approx 1.0$\,mag) between the stars and the
ring.

By comparing the azimuthal intensity variations of our models
to the observations, we have identified a range of possible dust
models at each wavelength assuming a simple power law size
distribution. While this likely is too simplistic an
assumption, it allows us to constrain the phase function
asymmetry parameters of the dust population at each
wavelength. Our best fit models have asymmetry parameters
$\overline{g_\lambda} \approx 0.4$--0.5 at all wavelength from
$I$ to $L'$. No ISM dust models are compatible with such a
behavior, nor are models representing the dust grain
population present in dense molecular clouds. The latter are
consistent with the $I$, $J$ and $H$ image of the ring (though
a pure power law size distribution is not) but not with the
new $L'$ image presented here. The only dust model that is
consistent with all images of the ring is that developed by
Wood et al. (2002) to reproduce the spectral energy
distribution of another protoplanetary disk. However, this
model predicts a 1\mic\ linear polarization rate that is much
smaller than what has been observed for the \ggt\ ring.

We therefore propose that the \ggt\ ring contains more or less
unprocessed molecular cloud-like dust grains in its surface
layers, $\sim50$\,AU above the midplane, while significantly
larger dust grains are present deeper into the disk, in
agreement with the millimeter thermal emission from the
system. We suggest that this stratified structure is a general
feature of protoplanetary disks and could be the result of
dust settling and/or preferred grain coagulation in the
deeper, densest parts of the disks. We also conclude that
large grains have grown in size in this large circumbinary
ring despite its much longer dynamical timescale with respect
to typical circumstellar disks where this has already been
suggested in the past.

%________________________________________________________________

\acknowledgements

We are grateful to the W. M. Keck Observatory personnel whose
professional and efficient help during our run allowed us to
obtain such high quality data as presented in this paper. In
particular, we thank Randy Campbell, Gary Puniwai, David le
Mignant, Peter Wizinowich and Tim Saloga. We also want to
thank John Krist for making his calibrated $HST$/WFPC2 $I$
band images of \ggt\ available to us for analysis. This work
has been supported by the National Science Foundation Science
and Technology Center for Adaptive Optics, managed by the
University of California at Santa Cruz under cooperative
agreement No. AST - 9876783, and by the Packard Foundation. Data
presented herein were obtained at the W.M. Keck Observatory, which
is operated as a scientific partnership among the California 
Institute of Technology, the University of California and the National
Aeronautics and Space Administration. The Observatory was made
possible by the generous financial support of the W.M. Keck
Foundation. The authors wish to recognize and acknowledge the
very significant cultural role and reverence that the summit
of Mauna Kea has always had within the indigenous Hawaiian
community. We are most fortunate to have the opportunity to
conduct observations from this mountain.

%________________________________________________________________

%________________________________________________________________
\clearpage
\begin{deluxetable}{cc}
\tablecaption{Properties of the \ggt\ binary system.}
\label{tab:binary}
\startdata
\tableline
Sep. & 0\farcs2507$\pm$0\farcs0015 \\
P.A. & 346\fdg0$\pm$1\fdg5 \\
$L'_A$ & 6.72$\pm$0.13 \\
$L'_B$ & 7.31$\pm$0.13 \\
$\Delta L'$ & 0.59$\pm$0.03 \\
\enddata
\tablecomments{The uncertainty in both components $L'$
magnitude include the uncertainty in the absolute photometric
calibration whereas the uncertainty on the flux ratio does
not.}
\end{deluxetable}

\begin{deluxetable}{lcccc}
\tablecaption{Observed properties of the \ggt\ ring.}
\label{tab:obs_disk}
\startdata
\tableline
Filter & $I$ & $J$ & $H$ & $L'$ \\
\tableline
$\lambda$ (\mic) & 0.81 & 1.00 & 1.55 & 3.78 \\
$R_\lambda$ (\%) & 1.30$\pm$0.03 & 1.40$\pm$0.03 &
1.54$\pm$0.03 & 0.97$\pm$0.09 \\ 
$FB_\lambda$ & 1.80$\pm$0.03 & 1.41$\pm$0.03 & 1.39$\pm$0.02 &
$\sim2.5$ \\
$D_{peak}$ (\arcsec) & 0.87$\pm$0.01 & 0.88$\pm$0.1 &
0.90$\pm$0.01 & 0.97$\pm$0.01 \\
$g_{fit}$ & 0.52$^{+0.12}_{-0.07}$ & 0.47$^{+0.19}_{-0.05}$ &
0.41$^{+0.14}_{-0.07}$ & 0.50$^{+0.27}_{-0.14}$ \\
\enddata
\tablecomments{$R_\lambda$ is the total disk/star flux ratio
whereas $FB_\lambda$ is the integrated front/back ratio for
the disk. $D_{peak}$ is the smallest distance (typically
measured around position angle 5--15\degr) between the center
of mass of the system and the peak intensity in radial
profiles of the ring. All of these quantities for the $J$ and
$H$ images are from McCabe et al. (2002) and those for $I$ band 
are from our reanalysis of the image obtained by Krist et al. 
(2002). Finally, $g_{fit}$
represents the asymmetry parameter of the best-fitting dust
model as defined in \S\,\ref{subsec:trends}; the associated
``uncertainties'' indicates the tolerable range of models.}
\end{deluxetable}

\begin{deluxetable}{clc|clc}
\tablecaption{List of all Monte Carlo simulations.}
\label{tab:models}
\startdata
\tableline
Model & $a_{max}$ & Filter & Model & $a_{max}$ & Filter \\
 & (\mic) & & & (\mic) & \\
\tableline
1 & 0.25 & I & 11 & 0.7 &   H \\
2 & 0.275 & I & 12 & 0.9 &  IJHL' \\
3 & 0.3 &  VIJHL' & 13 & 1.05 & L' \\
4 & 0.325 & J & 14 & 1.125 &  L' \\
5 & 0.35 & IJ & 15 & 1.25 &   IJHL' \\
6 & 0.4 &  IJH & 16 & 1.375 &  L' \\
7 & 0.45 &  IJH & 17 & 1.5 &    IJHL' \\
8 & 0.5 &   IJHL' & 18 & 2.0 &  L' \\
9 & 0.55 &  IH & 19 & 2.5 &   L' \\
10 & 0.6 &   H & 20 & 3.0 &  L' \\
\enddata
\end{deluxetable}

\begin{deluxetable}{ccccccc|ccccccc}
\tabletypesize{\scriptsize}
\tablecaption{Quantities extracted from the Monte Carlo
simulations.}
\label{tab:resmodels}
%\begin{tabular}{ccccccc|ccccccc}
\startdata
\tableline
$\lambda$ & Model & $a_{med}$ (\mic) & $\overline{g_\lambda}$ & $R_\lambda$ &
$FB_\lambda$ & $\chi^2$ & $\lambda$ & Model & $a_{med}$  (\mic) &
$\overline{g_\lambda}$ & $R_\lambda$ & $FB_\lambda$ &
$\chi^2$ \\
\tableline
$V$  & 3  & 0.205 & 0.73 & 0.0263 & 1.68 & \nodata & $H$  & 3  & 0.235 & 0.16 & 0.0501 & 0.36 & 73.8 \\
\cline{1-7}                              
$I$  & 1  & 0.184 & 0.38 & 0.0377 & 0.72 & 16.9 &     & 6  & 0.304 & 0.27 & 0.0402 & 0.58 & 26.7 \\
     & {\it 2}  & {\it 0.200} & {\it 0.45} & {\it 0.0363} & {\it 0.83} & {\it 11.7} &     & {\it 7}  & {\it 0.336} & {\it 0.34} & {\it 0.0377} & {\it 0.72} & {\it 18.2} \\
     & {\bf 3}  & {\bf 0.216} & {\bf 0.52} & {\bf 0.0339} & {\bf 0.93} & {\bf 11.3} &     & {\bf 8}  & {\bf 0.367} & {\bf 0.41} & {\bf 0.0357} & {\bf 0.87} & {\bf 14.6} \\
     & {\it 5}  & {\it 0.247} & {\it 0.64} & {\it 0.0306} & {\it 1.14} & {\it 12.9} &     & {\it 9}  & {\it 0.399} & {\it 0.48} & {\it 0.0339} & {\it 1.02} & {\it 18.9} \\
     & 6  & 0.278 & 0.71 & 0.0277 & 1.35 & 15.3 &     & {\it 10} & {\it 0.431} & {\it 0.55} & {\it 0.0323} & {\it 1.14} & {\it 18.7} \\
     & 7  & 0.307 & 0.73 & 0.0255 & 1.52 & 14.9 &     & 11 & 0.495 & 0.67 & 0.0286 & 1.37 & 22.3 \\
     & 8  & 0.335 & 0.75 & 0.0235 & 1.70 & 17.7 &     & 12 & 0.614 & 0.74 & 0.0232 & 1.75 & 30.8 \\
     & 9  & 0.362 & 0.78 & 0.0217 & 1.83 & 18.8 &     & 15 & 0.803 & 0.83 & 0.0180 & 2.16 & 33.6 \\
     & 12 & 0.529 & 0.87 & 0.0142 & 2.24 & 20.0 &     & 17 & 0.927 & 0.86 & 0.0153 & 2.26 & 32.3 \\
\cline{8-14}
     & 15 & 0.655 & 0.91 & 0.0114 & 2.33 & 20.8 &$L'$ & 3  & 0.242 & 0.003& 0.0745 & 0.25 & 44.2 \\
     & 17 & 0.719 & 0.92 & 0.0099 & 2.26 & 24.1 &     & 8  & 0.400 & 0.008& 0.0612 & 0.30 & 31.5 \\
\cline{1-7}                              
$J$  & 3  & 0.222 & 0.36 & 0.0387 & 0.74 & 17.2 &     & 12 & 0.692 & 0.24 & 0.0407 & 0.59 & 13.2 \\
     & {\it 4}  & {\it 0.238} & {\it 0.42} & {\it 0.0365} & {\it 0.82} & {\it 13.6} &     & 13 & 0.789 & 0.31 & 0.0372 & 0.77 & 8.52 \\
     & {\bf 5}  & {\bf 0.254} & {\bf 0.47} & {\bf 0.0359} & {\bf 0.99} & {\bf 13.0} &     & {\it 14} & {\it 0.837} &{\it 0.36} & {\it 0.0363} & {\it 0.87} & {\it 6.63} \\
     & {\it 6}  & {\it 0.285} & {\it 0.58} & {\it 0.0322} & {\it 1.08} & {\it 13.6} &     & {\it 15} & {\it 0.915} & {\it 0.43} & {\it 0.0344} & {\it 1.04} & {\it 6.70} \\
     & {\it 7}  & {\it 0.317} & {\it 0.66} & {\it 0.0297} & {\it 1.29} & {\it 15.1} &     & {\bf 16} & {\bf 0.995} & {\bf 0.50} & {\bf 0.0324} & {\bf 1.17} & {\bf 5.75} \\
     & 8  & 0.348 & 0.71 & 0.0272 & 1.43 & 17.6 &     & {\it 17} & {\it 1.076} & {\it 0.58} & {\it 0.0306} & {\it 1.35} & {\it 5.86} \\
     & 12 & 0.561 & 0.85 & 0.0170 & 2.10 & 23.3 &     & {\it 18} & {\it 1.387} & {\it 0.72} & {\it 0.0246} & {\it 1.74} & {\it 6.41} \\
     & 15 & 0.712 & 0.89 & 0.0129 & 2.33 & 23.7 &     & {\it 19} & {\it 1.670} & {\it 0.77} & {\it 0.0205} & {\it 2.04} & {\it 7.18} \\
     & 17 & 0.799 & 0.91 & 0.0113 & 2.50 & 24.2 &     & 20 & 1.937 & 0.82 & 0.0178 & 2.41 & 7.92 \\
\enddata
%\end{tabular}
\tablecomments{$\overline{g_\lambda}$ is the asymmetry
parameter of the phase function for the median scatterer,
whose size is $a_{med}$; this
is an input to the numerical simulations and not one of their
results. $\chi^2$ measures the quality of the fit to the
azimuthal intensity profile of the ring; the boldfaced entries
indicate the best-fit models while the italicized ones
represent acceptable models (see text for details).}
\end{deluxetable}

%________________________________________________________________

\pagebreak
\clearpage

\begin{figure}
%\plotone{f1.eps}
\vspace*{10cm}
\caption{{\it Left:} Final shift-and-add image of the \ggt\
system, displayed on a logarithmic scale. North is up and East
to the left. {\it Right:} PSF-subtracted image of the ring,
displayed on a square root stretch from the ring's brightest
point down to zero, after smoothing by median averaging pixels
in a sliding 3 pixel-radius aperture.}
\label{fig:images}
\end{figure}

\begin{figure*}
%\plotone{f4.eps}
\vspace*{10cm}
\caption{{\it Left:} PSF subtracted images of the \ggt\
ring at $I$ (Krist et al. 2002), $J$ and $H$ (McCabe et
al. 2002) and $L'$ (this work), from top to bottom. {\it
Right:} Synthetic intensity maps from the five Monte Carlo
models that were calculated at all wavelengths. The respective
values of $a_{max}$ are indicated in parentheses. All images
have been rotated so that North is up and East to the left and
are 4\farcs2$\times$3\farcs4 in size. The grayscale is linear
from zero to the peak of the ring in all cases.}
\label{fig:models}
\end{figure*}

\begin{figure*}
%\plotone{f5.eps}
\vspace*{10cm}
\caption{$H$ band images of the \ggt\ system emphasizing the
need for a revised geometry of the ring: {\bf a:} Observed
image (McCabe et al. 2002); {\bf b:} synthetic map for model 9
with $R_{in}=180$\,AU and a $1/e$ width of 7\,AU inside of
this radius; {\bf c:} same model with an artificial cutoff of
the ring 35\,AU above its midplane; {\bf d:} same model
without cutoff but with $R_{in}=190$\,AU and a interior $1/e$
width of only 2\,AU. The dashed lines, traced for guiding the
eye, represent the location of the peak of the ring along its
semi-minor axis (upper line) and the point furthest away from
the center of mass of the system (lower line). All images are
plotted on the same linear gray scale and
4\farcs2$\times$3\farcs4 size.}
\label{fig:comparmodels}
\end{figure*}

\begin{thebibliography}

\bibitem[1990]{beckwith90} Beckwith, S. V. W., Sargent, A. I.,
Chini, R. S. \& Guesten, R. 1990, AJ, 99, 924

\bibitem[1991]{beckwithsargent01} Beckwith, S. V. W. \&
Sargent, A. I. 1991, ApJ, 381, 250

\bibitem[1999]{bertout99} Bertout, C., Robichon, N. \& Arenou
F. 1999, A\&A, 352, 574

\bibitem[1991]{bouchet91} Bouchet, P., Schmider, F. X. \&
Manfroid, J. 1991, A\&AS, 91, 409

\bibitem[1996]{burrows96} Burrows, C. J., Stapelfeldt, K. R.,
Watson, A. M., Krist, J. E., Ballester, G. E., Clarke, J. T.,
Crisp, D., Gallagher, J. S. III, Griffiths, R. E.,  Hester,
J. J., Hoessel, J. G., Holtzman, J. A., Mould, J. R., Scowen,
P. A., Trauger, J. T. \& Westphal, J. A. 1996, ApJ, 473, 437

\bibitem[2001]{chiang01} Chiang, E. I., Joung, M. K.,
Creech-Eakman, M. J., Qi, C., Kessler, J. E., Blake, G. A. \&
van Dischoeck, E. F. 2001, ApJ, 547, 1077

\bibitem[2003]{clayton03} Clayton, G. C., Wolff, M. J., Sofia,
U. J., Gordon, K. D. \& Misselt, K. A. 2003, ApJ, 588, 871

\bibitem[2001]{cotera01} Cotera, A. S., Whitney, B. A., Young,
E., Wolff, M. J., Wood, K., Povich, M., Schneider, G., Rieke,
M. \& Thompson, R. 2001, ApJ, 556, 958

\bibitem[2001]{dalessio01} D'Alessio, P., Calvet, N. \&
Hartmann, L. 2001, ApJ, 553, 321

\bibitem[2000]{duchene00} Duch\^ene, G. 2000, Ph.D. thesis,
Univ. Grenoble

\bibitem[2003]{duchene03} Duch\^ene, G., M\'enard, F.,
Stapelfeldt, K. R. \& Duvert, G. 2003, A\&A, 400, 559

\bibitem[1993]{ghez} Ghez, A. M., Neugebauer, G. \& Matthews,
K. 1993, AJ, 106, 2005

\bibitem[1999]{guilloteau99} Guilloteau, S., Dutrey, A. \&
Simon , M. 1999, A\&A, 348, 570

\bibitem[2001]{haisch01} Haisch, K. E. Jr., Lada, E. A. \&
Lada, C. J. 2001, ApJ, 553, L153

\bibitem[2003]{hartigankenyon03} Hartigan, P. \& Kenyon,
S. J. 2003, ApJ, 583, 334

\bibitem[2002]{itoh02} Itoh, Y., Tamura, M., Hayashi, S. S.,
Oasa, Y., Fukagawa, M., Kaifu, N., Suto, H., Murakawa, K.,
Doi, Y., Ebizuka, N., Naoi, T., Takami, H., Takato, N.,
Gaessler, W., Kanzawa, T., Hayano, H., Kamata, Y. \&
Saint-Jacques, D. 2002, PASJ, 54, 963

\bibitem[1995]{kenyonhartmann95} Kenyon, S. J. \& Hartmann,
L. 1995, ApJS, 101, 117

\bibitem[2003]{kenyonbromley03} Kenyon, S. J. \& Bromley,
B. C. 2003, AJ, in press

\bibitem[1994]{kim94} Kim, S.-H., Martin, P. G. \& Hendry,
P. D. 1994, ApJ, 422, 164

\bibitem[2002]{krist02} Krist, J. E., Stapelfeldt, K. R. \&
Watson, A. M. 2002, ApJ, 570, 785

\bibitem[2003]{lachaume03} Lachaume, R., Malbet, F. \& Monin
J.-L. 2003, A\&A, 400, 185

\bibitem[1991]{leinert91} Leinert, C., Haas, M., Mundt, R.,
Richichi, A. \& Zinnecker, H. 1991, A\&A, 250, 407

\bibitem[2002]{mccabe02} McCabe, C., Duch\^ene, G. \& Ghez, A.
M. 2002, ApJ, 575, 974

\bibitem[2003]{mccabe03} McCabe, C., Duch\^ene, G. \& Ghez, A.
M. 2003, ApJ, 588, L113

\bibitem[1994]{manningsemerson94} Mannings, V. \& Emerson,
J. P. 1994, MNRAS, 267, 361

\bibitem[1989]{mathiswhiffen89} Mathis, J. S. \& Whiffen,
G. 1989, ApJ, 341, 808

\bibitem[1989]{menard89} M\'enard, F. 1989, PhD. thesis,
Univ. Montr\'eal 

\bibitem[1996]{pollack96} Pollack, J. B., Hubickyj, O.,
Bodenheimer, P., Lissauer, J. J., Podolak, M. \& Greenzweig,
Y. 1996, Icar., 124, 62

\bibitem[1992]{press92} Press, W. H., Teukolsky, S. A.,
Vetterling, W. T. \& Flannery, B. P. 1992, {\it Numerical
recipes in C} (2nd ed.), Cambridge Univ. Press

\bibitem[1996]{roddier96} Roddier, C. Roddier, F., Northcott,
M. J., Graves, J. E. \& Jim, K. 1996, ApJ, 463, 326

\bibitem[1969]{safranov69} Safranov, V. S. 1969, ``Evolution of
the Protoplanetary Cloud and Formation of the Earth and
Planets'' (Moscow: Nauka) (English transl. 1972, NASA TT
F-677)

\bibitem[2000]{silber00} Silber, J., Gledhill, T., Duch\^ene,
G. \& M\'enard, F. 2000, ApJ, 536, L89

\bibitem[1992]{simonguilloteau92} Simon, M. \& Guilloteau,
S. 1992, ApJ, 397, L47

\bibitem[1998]{stapelfeldt98} Stapelfeldt, K. R., Krist,
J. E., M\'enard, F., Bouvier, J., Padgett, D. L. \& Burrows,
C. J. 1998, ApJ, 502, L65

\bibitem[2001]{suttneryorke01} Suttner, G. \& Yorke,
H. W. 2001, ApJ, 551, 461

\bibitem[1993]{tinney93} Tinney, C. G., Mould, J. R. \& Reid,
I. N. 1993, AJ, 105, 1045

\bibitem[1980]{weidenschilling80} Weidenschilling, S. J. 1980,
Icar., 44, 172

\bibitem[2002]{weinberger02} Weinberger, A. J., Becklin,
E. E., Schneider, G., Chiang, E. I., Lowrance, P. J.,
Silverstone, M., Zuckerman, B., Hines, D. C. \& Smith,
B. A. 2002, ApJ, 566, 409

\bibitem[2001]{weingartnerdraine01} Weingartner, J. C. \&
Draine, B. T. 2001, ApJ, 548, 296

\bibitem[1996]{weis96} Weis, E. W. 1996, AJ, 112, 2300

\bibitem[1993]{wetherillstewart93} Wetherill, J. W. \&
Stewart, G. R. 1993, Icar., 106, 190

\bibitem[1999]{white99} White, R. J., Ghez, A. M., Reid, I. N.
\& Schultz, G. 1999, ApJ, 520, 811

\bibitem[2001]{whiteghez01} White, R. J. \& Ghez, A. M. 2001,
ApJ, 556, 265

\bibitem[2003]{whitney03} Whitney, B. A., Wood, K., Bjorkman,
J. E. \& Wolff, M. J. 2003, ApJ, 591, 1049

\bibitem[2000]{wizinowich00} Wizinowich, P., Acton, D. S.,
Shelton, C., Stomski, P., Dathright, J., Ho, K., Lupton, W.,
Tsubota, K., Lai, O., Max, C., Brase, J., An, J., Avicola, K.,
Olivier, S., Gavel, D., Macintosh, B., Ghez, A. \& Larkin,
J. 2000, SPIE, 112, 315

\bibitem[1999]{wood99} Wood, K., Crosas, M. \& Ghez, A. 1999,
ApJ, 516, 335

\bibitem[2002]{wood02} Wood, K., Wolff, M. J., Bjorkman, J. E.
\& Whitney, B. 2002, ApJ, 564, 887

\bibitem[1998]{zubko98} Zubko, V. G., Krelowski, J. \& Wegner,
W. 1998, MNRAS, 294, 548

\end{thebibliography}
\end{document}